\documentclass[12pt,a4paper]{article}
\usepackage{amsmath}
\usepackage{latexsym}
\makeatletter
\def\rddots{\mathinner{\mkern1mu\raise\p@%
    \vbox{\kern7\p@\hbox{.}}\mkern2mu%
    \raise4\p@\hbox{.}\mkern2mu\raise7\p@\hbox{.}\mkern1mu}}
\makeatother
\begin{document}

\title{\sl Comment on ``Epidemiological modeling of 
online social network dynamics"}
\author{
  Kazuyuki FUJII
  \thanks{E-mail address : fujii@yokohama-cu.ac.jp }\\
  International College of Arts and Sciences\\
  Yokohama City University\\
  Yokohama, 236--0027\\
  Japan\\
  }
\date{}
\maketitle
\begin{abstract}
  In this comment we give a simple analytic approximate 
solution to the infectious recovery SIR (irSIR) model 
given by J. Cannarella and J. A. Spechler \cite{CS}, which 
is a variant of the traditional SIR model.
\end{abstract}

\vspace{5mm}\noindent

\vspace{5mm}\noindent

\vspace{10mm}
Let us make a brief review of \cite{CS} within our necessity. 
The traditional SIR model (dynamics) is given by three 
variables $\{S=S(t), I=I(t), R=R(t)\}$ and a set of differential 
equations

\begin{eqnarray}
\label{eq:tr SIR}
\dot{S}&=&-\frac{\beta IS}{N} \nonumber \\
\dot{I}&=&\frac{\beta IS}{N}-{\gamma I} \\
\dot{R}&=&{\gamma I} \nonumber
\end{eqnarray}
where $\beta$ and $\gamma$ are constants. 
The meaning of variables S, I, R is respectively S (susceptible), 
I (infected) and R (recovered) and the population $N=S+I+R$ 
is independent of time. 

\vspace{3mm}
In order to fit from the spread of infections disease 
to the spread of online social networks 
Cannarella and Spechler changed (\ref{eq:tr SIR}) to be
\begin{eqnarray}
\label{eq:irSIR}
\dot{S}&=&-\frac{\beta IS}{N} \nonumber \\
\dot{I}&=&\frac{\beta IS}{N}-\frac{\nu IR}{N} \\
\dot{R}&=&\frac{\nu IR}{N} \nonumber
\end{eqnarray}
where $\beta$ and $\nu$ are new constants. They called 
this the infectious recovery SIR model (dynamics).

\vspace{3mm}
In the model there is no analytic solution (maybe), so 
they solved it {\bf at once} by use of a numerical method. 
Although it is not bad indeed we should study (\ref{eq:irSIR}) 
in detail from the mathematical point of view.

\vspace{3mm}
Let us rewrite (\ref{eq:irSIR}) like
\begin{eqnarray*}
\frac{\dot{S}}{S}&=&-\frac{\beta}{N}I \\
\frac{\dot{I}}{I}&=&\frac{\beta}{N}S-\frac{\nu}{N}R \\
\frac{\dot{R}}{R}&=&\frac{\nu}{N}I.
\end{eqnarray*}
Therefore, if we set
\[
S(t)=e^{s(t)},\ \ I(t)=e^{i(t)},\ \ R(t)=e^{r(t)}
\]
then (\ref{eq:irSIR}) becomes
\begin{eqnarray}
\label{eq:irSIR rewrited}
\dot{s}&=&-\frac{\beta}{N}e^{i} \nonumber \\
\dot{i}&=&\frac{\beta}{N}e^{s}-\frac{\nu}{N}e^{r} \\
\dot{r}&=&\frac{\nu}{N}e^{i}. \nonumber
\end{eqnarray}

\vspace{3mm}
Here, we use a simple approximation
\begin{equation}
\label{eq:simple approximation}
e^{x}=1+x
\end{equation}
in the following. Then (\ref{eq:irSIR rewrited}) becomes
\begin{eqnarray*}
\dot{s}&=&-\frac{\beta}{N}(1+i) \\
\dot{i}&=&\frac{\beta}{N}(1+s)-\frac{\nu}{N}(1+r) \\
\dot{r}&=&\frac{\nu}{N}(1+i)
\end{eqnarray*}
and we have the vector equation
\begin{equation}
\label{eq:vector equation}
\frac{d}{dt}
\left(
\begin{array}{c}
s \\
i \\
r
\end{array}
\right)
=
\frac{1}{N}
\left(
\begin{array}{c}
-\beta \\
\beta-\nu \\
\nu
\end{array}
\right)
+
\frac{1}{N}
\left(
\begin{array}{ccc}
0 & -\beta & 0 \\
\beta & 0 & -\nu \\
0 & \nu & 0
\end{array}
\right)
\left(
\begin{array}{c}
s \\
i \\
r
\end{array}
\right).
\end{equation}
Moreover, since
\[
\left(
\begin{array}{ccc}
0 & -\beta & 0 \\
\beta & 0 & -\nu \\
0 & \nu & 0
\end{array}
\right)
\left(
\begin{array}{c}
1 \\
1 \\
1
\end{array}
\right)
=
\left(
\begin{array}{c}
-\beta \\
\beta-\nu \\
\nu
\end{array}
\right)
\]
if we set
\begin{equation}
\label{eq:new variables}
\left(
\begin{array}{c}
\tilde{s} \\
\tilde{i} \\
\tilde{r}
\end{array}
\right)
=
\left(
\begin{array}{c}
s \\
i \\
r
\end{array}
\right)
+
\left(
\begin{array}{c}
1 \\
1 \\
1
\end{array}
\right)
\end{equation}
then the equation (\ref{eq:vector equation}) 
becomes a clear vector equation
\begin{equation}
\label{eq:final form}
\frac{d}{dt}
\left(
\begin{array}{c}
\tilde{s} \\
\tilde{i} \\
\tilde{r}
\end{array}
\right)
=
\frac{1}{N}
\left(
\begin{array}{ccc}
0 & -\beta & 0 \\
\beta & 0 & -\nu \\
0 & \nu & 0
\end{array}
\right)
\left(
\begin{array}{c}
\tilde{s} \\
\tilde{i} \\
\tilde{r}
\end{array}
\right).
\end{equation}

\vspace{3mm}
The solution is
\[
\left(
\begin{array}{c}
\tilde{s}(t) \\
\tilde{i}(t) \\
\tilde{r}(t)
\end{array}
\right)
=
\exp
\left\{
\frac{t}{N}
\left(
\begin{array}{ccc}
0 & -\beta & 0 \\
\beta & 0 & -\nu \\
0 & \nu & 0
\end{array}
\right)
\right\}
\left(
\begin{array}{c}
\tilde{s}(0) \\
\tilde{i}(0) \\
\tilde{r}(0)
\end{array}
\right)
\]
and from (\ref{eq:new variables}) we obtain
\begin{eqnarray}
\label{eq:result}
\left(
\begin{array}{c}
s(t) \\
i(t) \\
r(t)
\end{array}
\right)
&=&
\left[
\exp
\left\{
\frac{t}{N}
\left(
\begin{array}{ccc}
0 & -\beta & 0 \\
\beta & 0 & -\nu \\
0 & \nu & 0
\end{array}
\right)
\right\}
-
\left(
\begin{array}{ccc}
1 & 0 & 0 \\
0 & 1 & 0 \\
0 & 0 & 1
\end{array}
\right)
\right]
\left(
\begin{array}{c}
1 \\
1 \\
1
\end{array}
\right)+
\nonumber \\
&&\ 
\exp
\left\{
\frac{t}{N}
\left(
\begin{array}{ccc}
0 & -\beta & 0 \\
\beta & 0 & -\nu \\
0 & \nu & 0
\end{array}
\right)
\right\}
\left(
\begin{array}{c}
s(0) \\
i(0) \\
r(0)
\end{array}
\right).
\end{eqnarray}

\vspace{3mm}
The remaining task is to calculate the exponential. 
For the purpose let us list the following

\noindent
{\bf Formula}
\begin{eqnarray*}
&&
\exp
\left\{t
\left(
\begin{array}{ccc}
0 & -x & 0 \\
x & 0 & -y \\
0 & y & 0
\end{array}
\right)
\right\}
=   \frac{1}{x^{2}+y^{2}}\times  \\
&&
\left(
\begin{array}{ccc}
y^{2}+x^{2}\cos(t\sqrt{x^{2}+y^{2}}) & -x\sqrt{x^{2}+y^{2}}\sin(t\sqrt{x^{2}+y^{2}}) & xy(1-\cos(t\sqrt{x^{2}+y^{2}})) \\
x\sqrt{x^{2}+y^{2}}\sin(t\sqrt{x^{2}+y^{2}}) & (x^{2}+y^{2})\cos(t\sqrt{x^{2}+y^{2}}) & -y\sqrt{x^{2}+y^{2}}\sin(t\sqrt{x^{2}+y^{2}}) \\
xy(1-\cos(t\sqrt{x^{2}+y^{2}})) & y\sqrt{x^{2}+y^{2}}\sin(t\sqrt{x^{2}+y^{2}}) & x^{2}+y^{2}\cos(t\sqrt{x^{2}+y^{2}}) 
\end{array}
\right).
\end{eqnarray*}
The proof is a simple exercise of undergraduates, \cite{FO}, \cite{Five}.

By setting
\[
x=\frac{\beta}{N},\ \ y=\frac{\nu}{N}\ \ \mbox{and}\ \ \phi =\sqrt{\beta^{2}+\nu^{2}}
\]
the formula gives
\[
\exp
\left\{\frac{t}{N}
\left(
\begin{array}{ccc}
0 & -\beta & 0     \\
\beta & 0 & -\nu  \\
0 & \nu & 0
\end{array}
\right)
\right\}
=
\frac{1}{\phi^{2}}
\left(
\begin{array}{ccc}
\nu^{2}+\beta^{2}\cos(t\frac{\phi}{N}) & -\beta\phi\sin(t\frac{\phi}{N}) & \beta\nu(1-\cos(t\frac{\phi}{N})) \\
\beta\phi\sin(t\frac{\phi}{N}) & \phi^{2}\cos(t\frac{\phi}{N}) & -\nu\phi\sin(t\frac{\phi}{N}) \\
\beta\nu(1-\cos(t\frac{\phi}{N})) & \nu\phi\sin(t\frac{\phi}{N}) & \beta^{2}+\nu^{2}\cos(t\frac{\phi}{N})
\end{array}
\right)
\]
and from (\ref{eq:result}) it is easy to see the following
\begin{eqnarray}
\label{eq:main result}
s(t)
&=&
\frac{\beta\nu-\beta^{2}}{\phi^{2}}(1-\cos(t\frac{\phi}{N}))-\frac{\beta}{\phi}\sin(t\frac{\phi}{N})+ 
\nonumber \\
&&
\frac{\nu^{2}+\beta^{2}\cos(t\frac{\phi}{N})}{\phi^{2}}s(0)-
\frac{\beta\sin(t\frac{\phi}{N})}{\phi}i(0)+
\frac{\beta\nu(1-\cos(t\frac{\phi}{N}))}{\phi^{2}}r(0),  \nonumber \\
i(t)
&=&
\cos(t\frac{\phi}{N})-1+\frac{\beta-\nu}{\phi}\sin(t\frac{\phi}{N})+ \nonumber \\
&&
\frac{\beta\sin(t\frac{\phi}{N})}{\phi}s(0)+
\cos(t\frac{\phi}{N})i(0)-
\frac{\nu\sin(t\frac{\phi}{N})}{\phi}r(0), \\
r(t)
&=&
\frac{\beta\nu-\nu^{2}}{\phi^{2}}(1-\cos(t\frac{\phi}{N}))+\frac{\nu}{\phi}\sin(t\frac{\phi}{N})+ 
\nonumber \\
&&
\frac{\beta\nu(1-\cos(t\frac{\phi}{N}))}{\phi^{2}}s(0)+
\frac{\nu\sin(t\frac{\phi}{N})}{\phi}i(0)+
\frac{\beta^{2}+\nu^{2}\cos(t\frac{\phi}{N})}{\phi^{2}}r(0). \nonumber
\end{eqnarray}

\vspace{5mm}
Let us summarize our result.

\noindent
{\bf Result}\ \ Our analytic approximate solution to (\ref{eq:irSIR}) 
is given by
\begin{equation}
\label{eq:theorem}
S(t)=e^{s(t)},\ \ I(t)=e^{i(t)},\ \ R(t)=e^{r(t)}
\end{equation}
with (\ref{eq:main result}).

\vspace{2mm}
To check the validity of the approximate solution is left to 
some readers \footnote{The author doesn't have skill of fitting of 
curves in the numerical analysis}.

\vspace{5mm}
We conclude this note with one comment. Our approximation 
(\ref{eq:simple approximation}) ($e^{x}=1+x$) is too simple, 
so better one is 
\[
e^{x}=1+x+\frac{x^{2}}{2}.
\]
In general, there is no need to add further higher order terms 
from $\sum_{n=3}^{\infty}\frac{1}{n!}x^{n}$ because of 
non-linearity of the original equation. 
However, even in this case we don't know how to solve 
the equation
\begin{eqnarray*}
\dot{s}&=&-\frac{\beta}{N}\left(1+i+\frac{i^{2}}{2}\right) \\
\dot{i}&=&\frac{\beta}{N}\left(1+s+\frac{s^{2}}{2}\right)-
\frac{\nu}{N}\left(1+r+\frac{r^{2}}{2}\right) \\
\dot{r}&=&\frac{\nu}{N}\left(1+i+\frac{i^{2}}{2}\right)
\end{eqnarray*}
explicitly at the present time.

This is a challenging problem for young students.

\vspace{10mm}

\end{document}